\documentclass{acmtog}
\acmVolume{}
\acmNumber{}
\acmYear{}
\acmMonth{}
\acmArticleNum{}  
\acmdoi{}

\usepackage{amsmath}
\usepackage{gensymb}
\usepackage{algorithm}
\usepackage{algorithmic}
\usepackage[draft]{pgf}
\usepackage{comment}
\usepackage{wrapfig}
\usepackage{color}
\usepackage{multirow}
\usepackage{caption}

\usepackage{expl3}
\ExplSyntaxOn
\newcommand\latinabbrev[1]{
  \peek_meaning:NTF . {
    #1\@}%
  { \peek_catcode:NTF a {
      #1.\@ }%
    {#1.\@}}}
\ExplSyntaxOff

\def\etc{\latinabbrev{etc}}

\begin{document}

\markboth{L. Sun et al.}{Lens Factory: Automatic Lens Generation Using Off-the-shelf Components}

\title{Lens Factory: Automatic Lens Generation Using Off-the-shelf Components} 

\author{Libin Sun\footnotemark
\affil{Brown University}
\and 
Brian Guenter {\upshape and} Neel Joshi {\upshape and} Patrick Therien
\affil{Microsoft Research}
\and
James Hays
\affil{Brown University}}


\maketitle

\begin{figure*}[!t]
\includegraphics[width=1\textwidth]{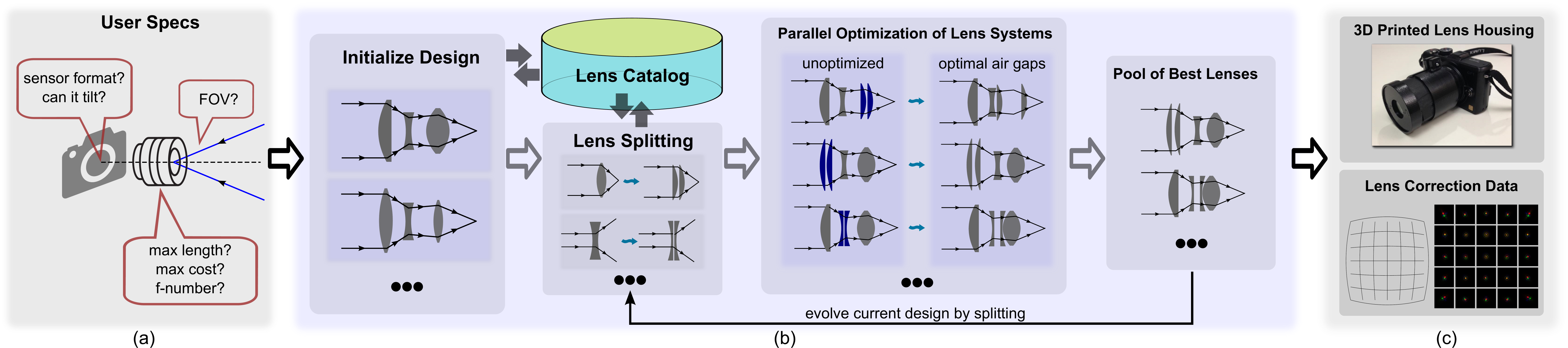}
\caption{A visualization of various components in our Lens Factory system. (a) The user provides specifications on the design through our UI. (b) Our system starts from a simple preset design and iteratively improves the design by substituting off-the-shelf lenses according to a set of splitting rules and continuously optimizing air gaps to maximize sharpness. (c) Finally, our system provides a 3D-printed lens housing for assembly and lens data for correcting distortion and lateral chromatic aberration.}
\label{fig:system_diagram}
\end{figure*}

\begin{abstract} 
Custom optics is a necessity for many imaging applications. Unfortunately, custom lens design is costly (thousands to tens of thousands of dollars), time consuming (10-12 weeks typical lead time), and requires specialized optics design expertise. By using only inexpensive, off-the-shelf lens components the \emph{Lens Factory} automatic design system greatly reduces cost and time. Design, ordering of parts, delivery, and assembly can be completed in a few days, at a cost in the low hundreds of dollars. Lens design constraints, such as focal length and field of view, are specified in terms familiar to the graphics community so no optics expertise is necessary. Unlike conventional lens design systems, which only use continuous optimization methods, \emph{Lens Factory} adds a discrete optimization stage. This stage searches the combinatorial space of possible combinations of lens elements to find novel designs, evolving simple canonical lens designs into more complex, better designs. Intelligent pruning rules make the combinatorial search feasible. We have designed and built several high performance optical systems which demonstrate the practicality of the system.
\end{abstract}

\footnotetext{The majority of the work is done while the first author was an intern at Microsoft Research.}

\section{Introduction}

Custom imaging systems can unlock powerful new capabilities in a variety of fields such as computer graphics, computer vision, computational photography, medical imaging, surveillance, virtual reality, and gaming~\cite{Wilburn:2005:HPI,Levoy:2006:LFM,Cossairt:2010,Pamplona:2010:NID,Brady:2012,Manakov:2013:RCA,Levin:2007:IDC:1275808.1276464,scalingLawForSphericalLenses,computationalCameras,ZhouFocalSweep}. These systems rely on the custom design of camera hardware including novel lens systems. 

Unfortunately, building a custom lens system is still the domain of optics experts. Modern lens design packages, such as Zemax and Code V, are expensive and have a steep learning curve for non optics experts. 

Even once these tools are mastered it is all too easy for a beginner or even an expert, to design a lens which cannot be manufactured. Understanding the physical properties of optical glasses and modern lens manufacturing processes is essential to success. This knowledge is difficult to acquire. Much of it is proprietary, poorly documented, or not documented at all, and acquired only through years of experience. For example, birefringence caused by stress in the plastic lens molding seriously degraded performance of the Aware2 gigapixel camera~\cite{Brady:2012}. Similar experiences in our own lab motivated us to build the Lens Factory system. Previously, we designed a lens with a sapphire element that had extraordinary performance. Fortunately, before manufacture we learned sapphire is birefringent which would have degraded performance tremendously. We contracted an optics company to design our next lens, but they accidentally designed a surface with curvature that couldn't be ground correctly by their equipment. This wasn't discovered until after the lenses were made, leading to poor performance. This is similar to perhaps the most famous lens manufacturing error -- the Hubble Space Telescope's main mirror, which was incorrectly ground and caused severe spherical aberration.

Even without these manufacturing difficulties the long lead time to build a lens -- 3 months or more -- slows the rate of research progress. If an error isn't discovered until the lens is built the delay and cost of another manufacturing cycle could easily cause project cancellation. This puts custom lens design out of the reach of all but the largest, most well-funded companies and university labs.




The \emph{Lens Factory} system dramatically reduces the cost and difficulty of custom lens design by automatically creating custom multi-element lens systems using off-the-shelf components. Other lens design packages, such as Zemax and Code V, are not designed to automatically create lens designs from scratch. They require significant user input and expertise to use. Lens Factory, by contrast, only requires the user to input a simple set of high level application specifications, such as the sensor size and desired field of view. Then our algorithm automatically explores the vast combinatorial search space of element choices. Design, ordering of parts, delivery, and assembly can be completed in a few days, at a cost in the low hundreds of dollars.

Lens Factory uses a combination of discrete and continuous optimization. Starting from a small number of simple lens design patterns, the system substitutes lens elements to generate a large number of candidates which are then evaluated using continuous optimization to set the air gaps between elements. To further improve performance, the system applies element splitting rules to introduce new lens component types and the discrete/continuous optimization is run again on the more complex system. After optimization is complete, a lens housing assembly is generated by 3D printing.

Our system initializes lens design using simple traditional lens configurations, such as the triplet and Double Gauss. However, the iterative splitting operations let us discover lenses that do not fall into known design categories. Our discovered lens systems are also likely to diverge from traditional designs because we independently optimize per-channel sharpness under the assumption that lateral chromatic aberration and other distortions can be fixed as a post-process by modern imaging systems (e.g. using methods such as~\cite{lensfitting_eccv2012}). Historically, lens designers would not have this freedom because of the constraint that all frequencies focus without distortion on a chemical medium (film), which does not permit non-trivial post-processing.

We show that our system is capable of designing effective novel lens systems for several interesting applications -- a standard lens for micro four thirds cameras, non-parallel projection view cameras, and head-mounted displays. Since Lens Factory is limited to off-the-shelf parts it is not a replacement for an expert lens designer or fully custom lenses. A custom design will always have better performance, because there will be many more degrees of freedom to optimize over. However, many optical applications do not justify the cost of a full custom design and, as we show with the lenses we have built, the performance of Lens Factory designs can be quite good.  


Lens Factory also does not currently design zoom lenses not because of any inherent theoretical difficulty with doing so but because zoom lenses require precise relative motion of lens elements, not just translation of the entire lens assembly. Current 3D printing is not up to the task of creating the smooth and precise cam shapes that are needed.

Lens Factory reduces the cost of custom optical design by a factor of thirty or more and fabrication time by a factor of twenty. This dramatic reduction in cost and time makes custom lens design practical for a much broader range of applications. While our optimization scheme appears complex, it is mostly hidden from the user, and the only user inputs required by our system are a few numbers that are well understood by non-experts.
To summarize, we compare and contrast traditional design process and our Lens Factory system in the following table Table \ref{tab:traditional_vs_ours}.

\begin{table}[h]
\tbl{Differences from Traditional Lens Design}{
\centering
\begin{tabular}{ r|c|c| }
\multicolumn{1}{r}{}
 &  \multicolumn{1}{c}{Traditional lens design}
 &  \multicolumn{1}{c}{Lens Factory} \\
\cline{2-3}
turnaround time  & months & days \\
\cline{2-3}
total cost & \$10,000s & \$100s \\
\cline{2-3}
skill level & optics expert & non-expert \\
\cline{2-3}
fabrication & might fail & verified \\
\cline{2-3}
\end{tabular}}
\label{tab:traditional_vs_ours}
\end{table}

We make the following contributions:
\begin{enumerate}
\item Lens Factory is the first system a non-expert can use to automatically create sophisticated multi-element optical systems.
\item We introduce effective continuous and discrete optimization strategies for selecting and positioning off-the-shelf lens components.
\item Lens Factory makes it possible to automatically generate specialty lenses at a fraction of the cost of consulting a lens designer, with fast turnaround time.
\end{enumerate}
\section{Related Work}
The majority of papers related to lens optimization deal with the continuous optimization problem. An initial candidate lens, usually designed by hand, is continuously optimized to improve its performance. The number of elements and the glass types are chosen by the designer and fixed during the optimization; only the surface shapes and element separations are varied. Because the shapes of lens elements are modified during optimization it is often expensive to fabricate such a lens system. Off-the-shelf components could not be used as in our system.

Typical objective functions include minimizing spot size or optical path difference (OPD), or maximizing the MTF response at desired frequencies. Spot size or OPD optimization usually cannot yield maximum MTF performance, but MTF optimization early in the design process can fail to converge if the initial lens design has poor optical performance \cite{smith_MOE,smith_lensdesign}.

A common strategy is to first optimize spot size or OPD and then switch to MTF optimization. The recent work of Bates \shortcite{BatesMTF:doi:10.1117/12.868932} describes a method which uses through focus MTF as the objective function. This avoids the convergence problems associated with early use of an MTF objective function and uses only one function, rather than requiring a manual switch between different objectives.

Commercial systems such as Zemax have a feature which will replace any element in a system with the closest matching stock element. This requires the user to start with a fully optimized lens design and does not address the issue of choosing the best possible combination of stock elements, nor does it automatically split lens elements to evolve higher performance lenses. 

The most closely related system to ours is Cheng et al. \shortcite{RapidStock:doi:10.1117/12.2075390}. Starting from an existing lens design, they use Code V macros to replace each element in the lens with a single stock lens and then re-optimize. If a single element replacement is insufficient they try replacing with a cemented 2-element plano-convex or plano-concave lens. The disadvantages of this system are: (1) it only works with Code V, an expensive proprietary lens design system, (2) it requires a high quality existing lens design as a starting point, and (3) it does not apply a general set of lens splitting rules to generate new candidate lenses.

Less closely related work \cite{AutoAdditionCheng:03} uses lens-form parameters to determine whether, and where, to insert a new lens element or whether to delete an existing one. However there is no attempt to find a good stock lens candidate for the new lens element. Instead the newly inserted element is continuously optimized into the best form, which might be expensive to fabricate. 

A system for doing stock lens substitution for high power laser applications is described in Traub et al. \shortcite{StockLasers:doi:10.1117/12.2074508}. They use the ZPL macro programming language to take an existing lens design and substitute a single stock lens for each original lens element. Since lasers are monochromatic, they do not address the issue of minimizing chromatic aberration. Each original lens is converted to a double convex form if the original is roughly symmetric, or a plano-convex or plano-concave otherwise. The system does not create designs from scratch. An experienced lens designer must create the initial lens design and figure out how to make this design meet high level system specifications. In addition the system runs on the proprietary and expensive Zemax optical design program.
\begin{figure}[!t]
\begin{center}
   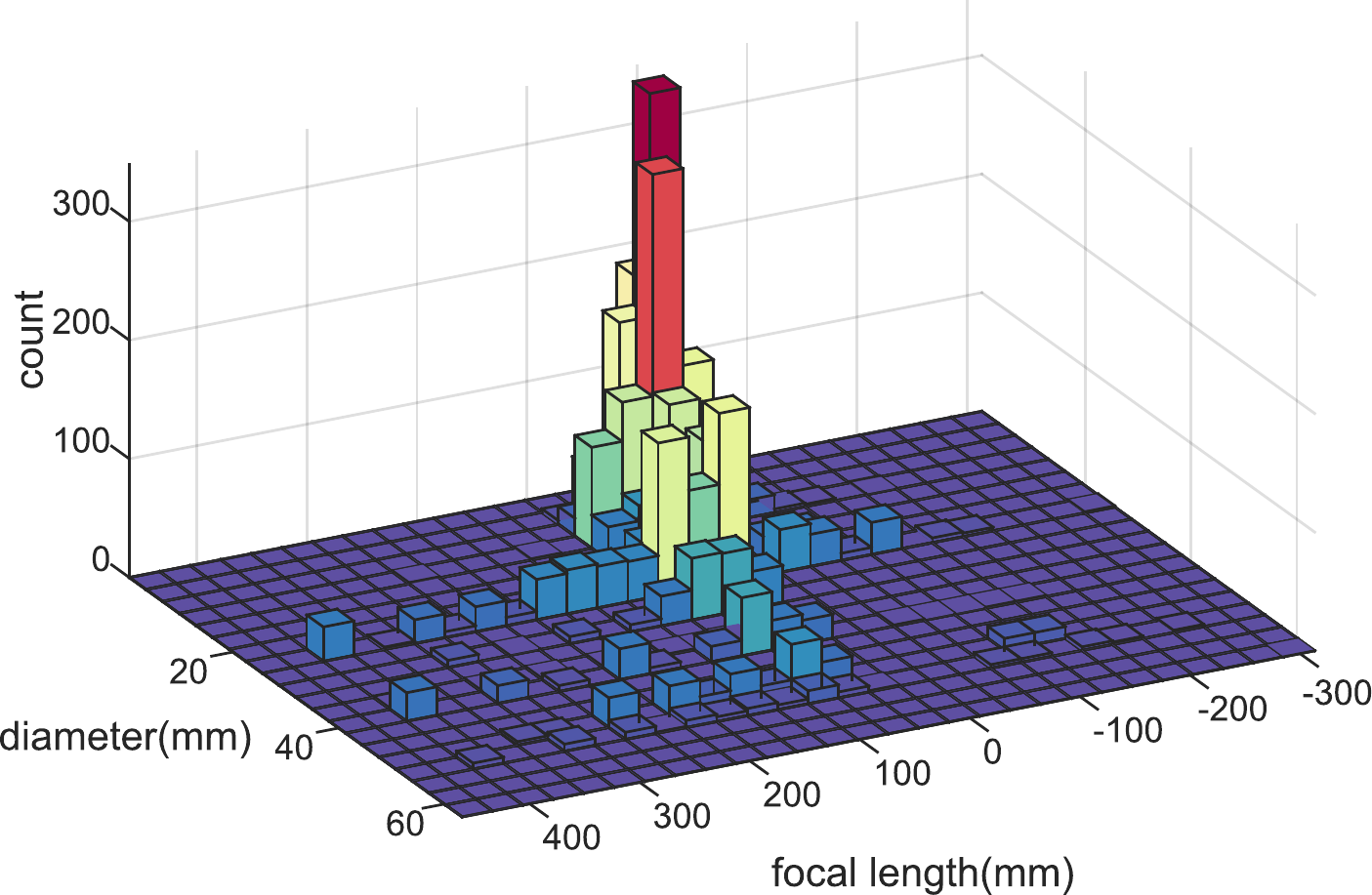
\end{center}
\vspace{-3pt}
\caption{2D histogram of diameters and focal lengths available in our lens catalog. Most lenses are smaller than 30mm in diameter.}
\label{fig:catalog}
\end{figure}
\section{Lens Factory System}
\label{sec:oursystem}

Designing a lens with Lens Factory begins with setting up camera and lens system specifications via our user interface system (see Figure \ref{fig:UIimg}). The user can specify object plane distance, field of view (FOV), f-number, camera body format, and optional optimization parameters. Any change in parameters is reflected via a system sketch in real-time.

The UI then saves the data files and scripts to run the optimization over a cluster of machines. Our lens optimization alternates between two phases: discrete search and continuous optimization. In the discrete phase lens elements are chosen from the catalog to satisfy the specifications and the physical constraints. In the continuous phase the air gaps between the elements are optimized to maximize system sharpness as measured by the spot size or the Modulation Transfer Function (MTF), or a combination of both.

The discrete element search space is too large to search exhaustively so a variety of pruning strategies are employed to make the computation feasible (see Section \ref{sec:implementation}). The system initializes the design from known lens forms such as the triplet and the Double Gauss and improves the design using a set of element splitting rules to introduce new lens components into the system (see Section \ref{sec:lenssplitting}). Our system is also capable of conducting a Monte Carlo based tolerance analysis to account for potential inaccuracies in the lens assembly, and allow the user to select a desired lens system from the top performing candidates.


Finally a 3D printed lens housing is made and the selected elements are snapped in place. The system can optionally generate a calibration file for correcting lateral chromatic aberration and distortion as a post-process.

\subsection{Off-the-shelf Lens Catalog}
The vendors Edmund Optics, Newport, Comar, and Thorlabs document their lenses precisely enough to be used in an optical design system. We collected a total of 3924 lens specifications from their websites. 
These lenses are spherical lens forms such as double-convex (DCX), double-concave (DCV), plano-convex (PCX), plano-concave (PCV), achromats (Ach-Pos and Ach-Neg), as well as a limited collection of meniscus lenses. 88\% of these lenses have positive power.

From the website information we generated an element catalog which contains the focal lengths, radii, center thicknesses, diameter, glass types, cost, and anti-reflection coating material for every lens element. After merging elements which differ only in anti-reflection coating, there are 770 positive lenses and 115 negative lenses.

One challenge to our System is that these components are a very limited and discrete sampling of the continuous lens parameter space. Figure \ref{fig:catalog} shows a 2D histogram of element diameter vs. focal length. The distribution is strongly peaked for lenses less than 30mm in diameter. Due to the limited number of meniscus lenses, there are few choices of bending, which is an important design axis for reducing aberrations.

\begin{figure}[t]
\begin{center}
   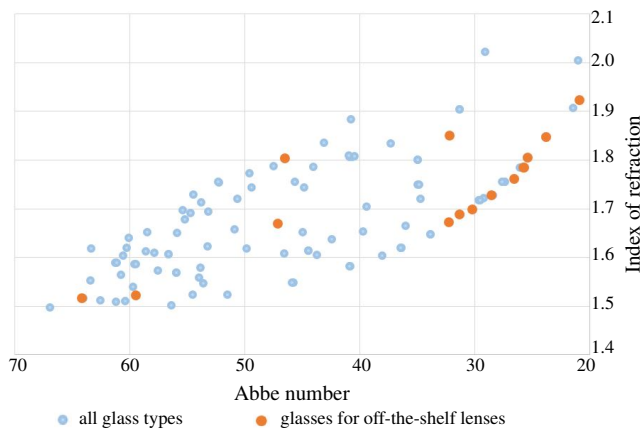
\end{center}
\vspace{-3pt}
\caption{A glass map by index of refraction vs. Abbe number, which is a measure of light dispersion. The smaller the Abbe number the more the index of refraction varies as a function of light wavelength. The glass types available for off-the-shelf lenses (orange) sample the glass map very sparsely.}
\label{fig:glasses}
\end{figure}

As shown in Figure \ref{fig:glasses}, glass choice is also limited. Typical commercially available glasses are shown in blue, while the glasses from our catalog are shown in orange. Only a tiny fraction of the glass space is available off-the-shelf. 

Since degree of bending and glass choice are two of the most important degrees of freedom for correcting aberrations, our lens design task is very different from traditional lens optimization. 

\subsection{User Interface}
\label{sec:UI}

\begin{figure}[t]
\begin{center}
   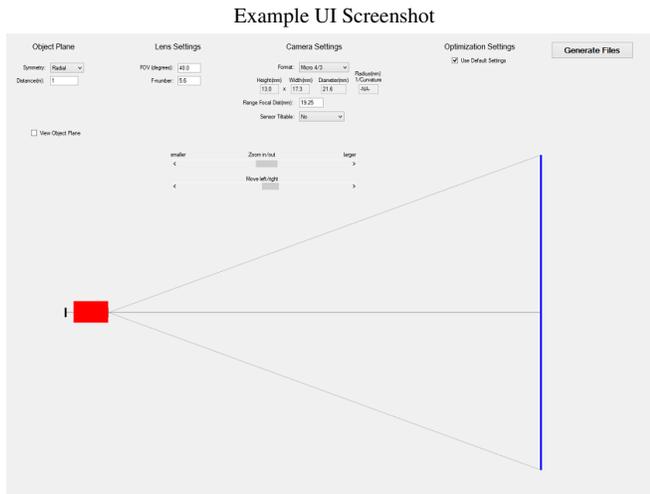
\end{center}
\vspace{-4pt}
\caption{Our user interface: the example screenshot captures a user designing a standard micro 4/3 camera lens. More details can be found in the supplementary video.}
\label{fig:UIimg}
\end{figure}

As shown in our supplementary video, our UI is intuitive and simple to use and provides a real-time sketch of the imaging setup as the user interacts. There are four main groups of settings that control properties of (1) the object plane, (2) the lens system, (3) the camera body, and (4) optional optimization parameters. A sample UI screenshot is shown in Figure \ref{fig:UIimg}. Our system is flexible enough to handle custom defined sensor formats (planar or curved), as well as tilted object plane for non-parallel projection. In the optional settings, the user can also specify the maximum number of lens elements to use, the maximum dimensions of the system, as well as total budget in dollars, \etc. Our supplementary video highlights the user interaction in designing a standard micro 4/3 lens and a view camera lens.

To illustrate our optimization procedure in subsequent sections, we will use a micro 4/3 lens with FOV 40\degree (or 30mm focal length) at f5.6 as a running example.

\subsection{Initializing a Design}
\label{sec:init}
A brute-force search over a multi-element system quickly becomes infeasible. Fortunately, hundreds of years of lens design expertise provides us with a few well studied classic lens forms that we use as a starting point to greatly reduce the search space. 

Our system begins the discrete optimization with simple existing lens design forms such as the triplet and the Double Gauss. Elements in the starting design are replaced with elements from the catalog that are of the same type (positive or negative power), but not necessarily the same focal length or diameter. Hundreds of thousands of candidate lenses may be tested in this phase but only a subset of these is passed on to the continuous optimization phase.  


As an example let's begin the micro 4/3 design with a triplet form, which consists of two positive outer elements and a negative middle element. An exhaustive search would examine $770\times115\times770\times2=136$ million combinations (considering 2 possible stop positions). If we consider the two possible orientations for asymmetric lens elements, the space is much greater. Finding the optimal air gaps for all these lens systems is clearly infeasible.

By constraining each element's power and diameter to be within $25\%$ of the value of a particular base triplet design from the search space shrinks to $23\times6\times43\times2=11868$ combinations, 11 thousand times fewer possibilities. We further prune the search space by quickly testing for requirements on the desired FOV, flange focal distance, \etc, which avoids the expensive continuous optimization step $90\%$ of the time. More details can be found in Section~\ref{sec:implementation}.

\begin{figure}[t]
\begin{center}
   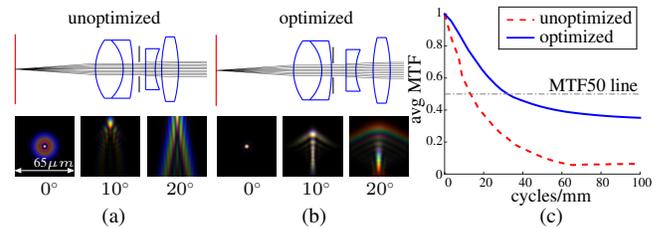
\end{center}
\vspace{-4pt}
   \caption{Visualization of a triplet design before and after the continuous optimization process. PSF's are shown for $0\degree$,$10\degree$ and $20\degree$. (a) air gaps between lens elements are set to 2mm before optimization. (b) After our two-stage continuous optimization, the center PSF appears significantly more peaked and the corner PSF's exhibit less aberrations leading to better image sharpness. (c) Significant improvement in MTF performance is observed after the optimization (higher is better). MTF50 response has approximately doubled.}
   \label{fig:continuous_opt}
\end{figure}

\subsection{Continuous Optimization for Air Gaps}
\label{sec:continuous_opt}
The continuous optimization itself has two phases. Ultimately we desire to maximize MTF, because this is strongly correlated with perceived image quality. But MTF optimization is prone to being trapped in local minima if applied early in the optimization when the lens performance is poor \cite{smith_MOE,smith_lensdesign}.

Minimizing spot size, or optical path difference (OPD) is less prone to being trapped in local minima, but does not give the best MTF response. We first minimize spot size and then maximize MTF, which has proven to be relatively immune to local minima in many existing lens design tools.

Our lens system  $\mathcal{L}^k_{\mathbf{c},\mathbf{d}}$ is a fixed sequence of $k$ optical elements $\mathbf{c}=[c_1, c_2, c_3,...,c_k]$ and the air spaces $\mathbf{d}=[d_1, d_2,...,d_{k-1}]$ between adjacent elements. Each $c_i$ is either a lens from our lens catalog or a stop. Air gaps are non-negative to avoid interpenetration of lens elements. 

\vspace{4pt}
 
\textbf{Optimizing for Sensor Air Gap.} Given a fixed lens system configuration $\mathcal{L}^k_{\mathbf{c},\mathbf{d}}$ and a well-defined objective function $\mathcal{F}$ that measures the imaged sharpness of point light sources $\{e_i\}$, we seek to find an optimal back focal length (BFL) $d_k^*=\arg\min\mathcal{F}(\mathcal{L}^k_{\mathbf{c},\mathbf{d}},  d_k)$, where $d_k$ is the air gap between the sensor and the last optical surface in the system. This is similar to the auto-focus mechanism in digital cameras. In particular, rays of wavelength $\lambda$ from each point light source $e_i$ in the object plane are traced through the lens system and land on the sensor. Our objective summarizes statistics from the these rays as follows:
\begin{equation}
\mathcal{F}(\mathcal{L}^k_{\mathbf{c},\mathbf{d}}, d_k) = \frac{1}{3n}\sum_{j=1}^3\sum_i^n f(e_i, \boldsymbol{\lambda}_j, \mathcal{L}^k_{\mathbf{c},\mathbf{d}}, d_k)
\label{eq:objective}
\end{equation}
where $j$ indexes through different color channels, $i$ indexes through the sampled emitter positions, $f$ is a function measuring spot size or OPD via geometric ray tracing, and $\boldsymbol{\lambda}_j$ is a set of representative wavelengths for the $j^{th}$ color channel. For spot size, we compute the MSSE w.r.t. the centroid of the spot diagram. For OPD, we compute the MSSE w.r.t. the mean optical path. 

As a deliberate design choice, we do not minimize the combined spot size of the red, green, and blue channels. Instead, the spot size from each channel is independently computed and then summed in Eq.\ref{eq:objective}. This allows the lens system to have small amounts of lateral chromatic aberration, which are easily corrected as a post process.

The sensor position is initialized by tracing a single paraxial ray close to the optical axis from a point light source at infinity. The sensor is placed at the intersection of this ray with the optical axis, if an intersection exists. Then $\mathcal{F}$ is minimized w.r.t. $d_k$ via gradient descent. Derivative computation uses finite differences with Richardson extrapolation to the limit.

\vspace{4pt}

\textbf{Optimizing for Lens Air Gaps.}
The ultimate goal is to optimize for $\mathbf{d}$ given $\mathbf{c}$, namely, to pick a set of air gap values $\mathbf{d^*}=\arg\min\mathcal{F}(\mathcal{L}^k_{\mathbf{c},\mathbf{d^*}}, d_k^*)$, where $d_k^*$ is the optimal BFL recomputed as described above for any given inter lens air gap configuration. We use gradient descent to optimize $\mathbf{d}$ but attempt to break out of any local minima with a local search. In particular, we conduct local search by grid search with small discrete steps for each air gap around its current value. We initialize the optimization by placing all optical elements equal distance apart, setting $\mathbf{d}=[a,a,...,a]$, and testing for a fixed set of values for $a \in \{1,2,...,6 \text{mm}\}$. We pick the best $a$ value (lowest objective cost) for initializing the gradient descent step.

\vspace{4pt}
\textbf{Second Stage Optimization.} After spot size optimization has converged, we replace $f$ with a function that measures MTF performance. Since geometric ray tracing cannot account diffraction effects, we render the Point Spread Function (PSF) via wave optics simulation using the Rayleigh-Sommerfield diffraction integral, then compute the area under the MTF curves, which is obtained by taking the Fourier Transform on the PSF. 

As can be seen in Figure \ref{fig:continuous_opt}, the triplet system for the standard micro 4/3 lens has significantly higher MTF response after the continuous optimization phase.

\subsection{Discrete Optimization through Lens Splitting}
\label{sec:lenssplitting}
The quality of lens systems can often be improved by adding additional lenses. For example, commercial SLR lenses commonly have six to ten elements. Variable zoom lenses, which we do not address, often have upwards of 15 components. Adding more lenses has two drawbacks, though --  the fabrication cost increases and the design space becomes combinatorially larger. We limit cost by letting users specify the maximum number of elements and total budget cost in the user interface (see Section~\ref{sec:UI}), and we deal with the large search space with several heuristics described below.

Given a lens design (e.g. an intermediate result in our optimization), simply adding a random lens will change the overall system power which changes effective focal length, and violates user specifications such as the field of view. Instead we \textit{split} an existing lens element into two elements and re-optimize. Distributing the power of a single lens over two elements reduces element curvature, which in turns reduces spherical aberration. Lens systems with lower mean squared refractive power tend to perform better~\cite{Sasian_Descour,AutoAdditionCheng:03}.
 With careful choice of power distribution and air gap size between the new elements, the overall power of the system will stay unchanged, delivering the same FOV while improving imaging quality.

\begin{figure*}[t]
\includegraphics[width=1\textwidth]{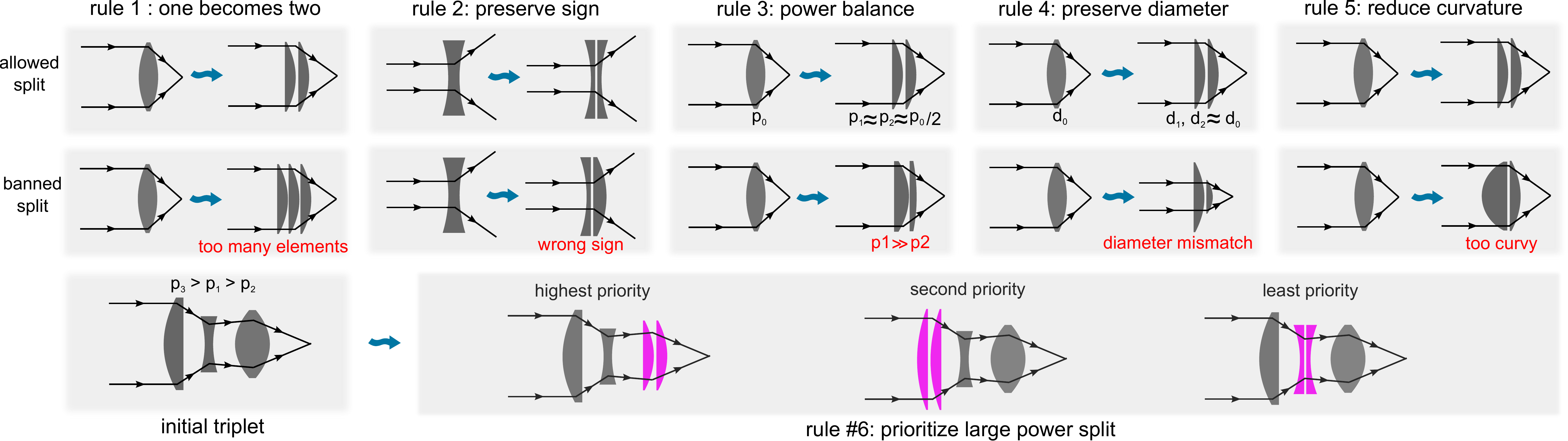}
\caption{A visualization of our splitting rules. Rule 1 through 5 define the basic properties a single split must follow. These rules are to be applied to any lens element in a multi-element design. Rule 6 puts other rules into context by strategically selecting an element for splitting that is most conducive for discovering better lens systems.}
\label{fig:rules}
\end{figure*}

We use the following one-to-two lens splitting rules, visualized in Figure \ref{fig:rules}:
\begin{enumerate}
\item[1] Splitting from one element to two: lens $l_0$ of power $p_0$ can be split into $l_1$ with power $p_1$ plus $l_2$ with power $p_2$. We do not consider more complex substitutions (e.g. one lens splits into three) because that would enormously increase the search space.
\item[2] Splitting should only use lenses with the same sign of power: $sign(p_0)=sign(p_1)=sign(p_2)$. For example, a negative element can only be replaced with two negative lenses.
\item[3] Splitting should lead to approximately equal distribution of powers: $(1-\alpha)*|p_0|/2<|p_i|<(1+\alpha)*|p_0|/2, i\in\{1,2\}$, where $0<\alpha<1$. Larger $\alpha$ allows for more extreme power combinations to be considered but can lead to a prohibitively large search space. We set $\alpha=0.25$.
\item[4] Splitting should preserve diameter of elements: diameters of $l_1$ and $l_2$ should be within $\pm25\%$ of diameter of $l_0$. This constraint reduces the occurrence vignetting as the lens system gets longer and more complex.
\item[5] Splitting should reduce maximum curvature: $l_1$ and $l_2$'s maximum curvature should be no larger than that of $l_0$'s. This constraint helps reduce aberrations (especially towards the corners) after splitting.
\item[6] Splitting should preferably occur where refractive power is concentrated \cite{AutoAdditionCheng:03}, hence placing priority on splitting lens elements with large curvatures and high power.
\end{enumerate}

Splitting can be carried out repeatedly to iteratively improve a design until performance converges or the maximum number of elements is reached. After each splitting operation, we test all possible positions of the stop. For instance, a lens system split into 4 elements would instantiate 5 continuous optimization tasks, one for each possible stop position.

\begin{figure}[t]
\begin{center}
   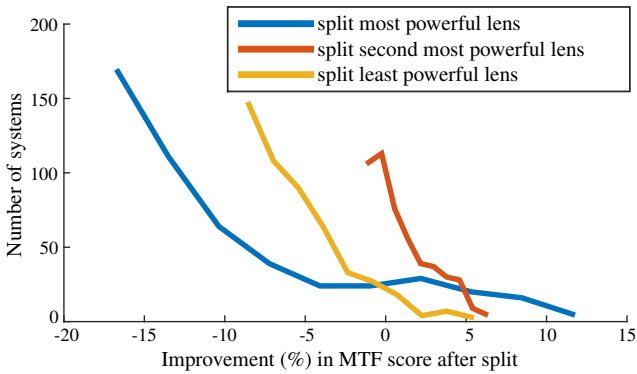
\end{center}
\caption{Splitting elements with higher power has a more significant effect on lens system performance and leads to the discovery of both worse and better designs. The superior  configurations (the long tail to the right of the plot) \emph{cannot} be discovered by splitting weaker elements, so given a fixed computational budget it is more favorable to prioritize splitting the most powerful elements in the system to maximize performance gain.}
\label{fig:splitrank_vis}
\end{figure}

\vspace{4pt}
\textbf{Effect of Splitting Powerful Elements.}
We bias our search towards splitting the most powerful optical elements of the lens system. We use a greedy selection criteria to rank lens elements by their maximum power and prioritize splitting according to this ranking. As shown in Figure \ref{fig:splitrank_vis}, splitting elements of higher power allows the system to discover high performing configurations (represented by the longer tail in the plot) which cannot be discovered by splitting less powerful elements, at the cost of having larger variance. Still, we can expect the performance upper bound to increase by splitting the more powerful elements.

\vspace{4pt}
\textbf{Evolving a Design by Continued Splitting.} A single round of splitting increases the number of elements in the system by one. To obtain a $k$-element design, we could start from the triplet and split $k-3$ times consecutively. Here we compare four evolution strategies to carry out multiple rounds of splitting:
\begin{enumerate}
\item[1] Random: all components are independently selected from the catalog at random.
\item[2] Greedy evolution: after each round of splitting, we take the single best lens configuration as the only starting point for splitting in the next round.
\item[3] Pooled evolution: after each round of splitting, we keep the top $n$ lens configurations (ranked by area under MTF curves). The group of $n$ candidates at iteration $t$ is $\mathbf{c}^t$. The next round of splitting uniformly samples among these $n$ candidates at random and applies our splitting procedure to form $\mathbf{c}^{t+1}$. We set $n$ to 60.
\item[4] Pooled evolution with swap: same as 3, except that the top $n$ lens systems are allowed to swap elements before a split takes place. Much like a mutation operator in genetic algorithms, each of the $k$ slots in a $k$-element system has $n$ possible candidates, $\mathbf{c}^t$ is formed by picking one (out of $n$) candidate per slot at random, followed by procedure 3 to generate $\mathbf{c}^{t+1}$.
\end{enumerate}

\begin{figure}[t]
\begin{center}
   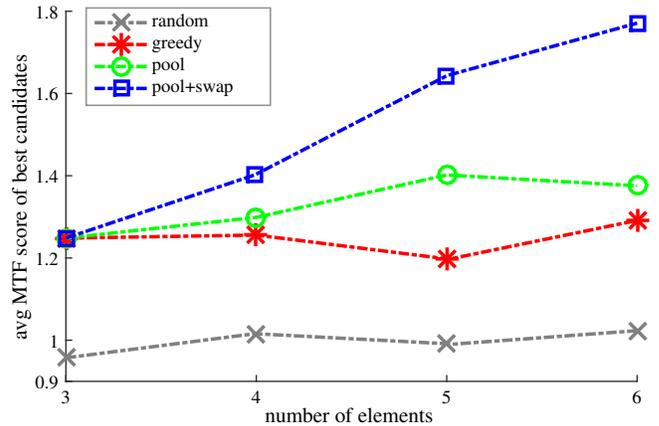
\end{center}
\vspace{-5pt}
\caption{Comparison of various evolution strategies using the standard micro 4/3 lens as an example. Pool+swap is able to steadily increase performance upper bound by keeping a diverse set of candidates to evolve on and allowing candidates to mutate via swapping elements, whereas other strategies are challenged by local minima.}
\label{fig:evolution_micro43}
\end{figure}

We compare these strategies by evolving from a triplet standard micro 4/3 design, and show performance progression across iterations in Figure \ref{fig:evolution_micro43}. We measure performance by the average MTF score of the best lens candidate found under each strategy. For each data point in Figure \ref{fig:evolution_micro43}, we assign a total budget of 1200 CPU hours over a cluster of 600 nodes for each strategy to explore the search space to ensure fair comparison.
For strategy 2, 3 and 4, the initial triplet designs are shared, as detailed in Section \ref{sec:init}, so their starting point MTF scores are identical. 

As expected, \emph{random} performs the worst because it draws independent lens elements at random, without considering promising candidate configurations from previous iterations. The \emph{greedy} strategy only makes use of a single lens candidate, disregarding other potentially useful configurations, and hence has a tendency to get stuck in local minima. The \emph{pool} strategy keeps a diverse set of top performing candidates at each split iteration to avoid local minima and is able to deliver a more steady increase in performance. 

However, to better avoid local minima and expand the search space more intelligently, the \emph{pool+swap} strategy is superior at discovering high performing lenses. Our experiments show a steady increase in performance upper bound without plateau. More details on the performance of the best lens discovered here can be found in Section \ref{sec:results_micro43}.

\subsection{Tolerance Analysis}
\label{sec:tolerance}
Any physical fabrication procedure cannot exactly match the optimized parameters of our lens systems. There are two main sources of errors in the fabricated lens: (1) errors due to slight deviation from the vendor supplied specs in diameter, thickness, curvatures, glass properties, \etc. (2) errors due to slight misalignment in the lens housing assembly, errors in actual air gaps, \etc. Since the first source of error is beyond our control, we focus on the second type instead.

We would like to verify that our discovered lens systems will perform well in light of these fabrication errors.  We conduct tolerance analysis by introducing random perturbations to lens system parameters. The magnitude of perturbations is based on what we can expect from modern 3D printing technology.


We conduct a Monte Carlo simulation by introducing i.i.d. random perturbations $\delta_i$ to each air gap $d_i$ by drawing from a Gaussian distribution $\delta_i\sim N(0, \sigma^2)$. We also impose a maximum perturbation amount $T$ such that $\delta_i\leq T$. Such perturbations include translation along the optical axis and decentering of the lens parts (including the stop) off the optical axis. The sensor is allowed to refocus after all perturbations to the lens parts have been done, and a final random perturbation is introduced to translate the sensor, simulating the error in mounting the lens assembly onto the camera body. We use $\sigma=20\mu m$, $T=100\mu m$ for perturbing the lens parts and stop, $\sigma=100\mu m$, $T = 300\mu m$ for perturbing the sensor.

We compute the MTF performance for each independent run in the Monte Carlo simulation, and tally the results to represent an estimated performance range to expect for the actual fabricated lens. Typically 20000 independent runs are used for a single design. Since this step is computationally expensive, the user can selectively conduct the tolerance analysis only for the top performing designs reported by Lens Factory at the end of the iterative splitting process. We report and compare expected performance against measured values in Section \ref{sec:results}.

\subsection{Implementation and Speedup}
\label{sec:implementation}

\textbf{Quick Pruning Test} Our catalog allows a vast space of possible lens configurations for most designs. Allowing the system to quickly \emph{fail} a given configuration is key to faster exploration of the space. This is achieved by a set of quick tests to decide if the expensive continuous optimization step should be carried out.
\begin{enumerate}
\item [1] \emph{Basic test}: checks if the lens cost is within user specified budget, and if the system dimensions are within user specified limits.
\item [2] \emph{Focus test}: the system shoots a single ray near the optical axis and test to see if it intersects the optical axis in the accepted range behind the last optical surface of the lens. This is defined by the flange focal distance in the UI. Afocal systems are never considered.
\item [3] \emph{f-number test}: the system tests to see if the target f-number can be achieved by adjusting the aperture size. 
\item [4] \emph{FOV test}: the system checks if the actual FOV of the lens system is close to the desired value specified by the user.
\item [5] \emph{Vignetting test}: the system tracks measured luminance on the sensor across the field to see if the desired luminance fall-off is met.
\end{enumerate}

\vspace{4pt}
\textbf{Caching Exit Rays.} While our system is capable of fast ray tracing, the number of rays increases linearly with the number of surfaces in the system, and each extra lens elements introduces at least 2 surfaces. A careful examination of our continuous optimization process (\ref{sec:continuous_opt}) reveals that we can speed up the inner loop of optimizing $d_k$ by caching all the rays that exit the last surface of the lens system. This is because all optical surfaces in front of the sensor and their airgaps are fixed during the optimization of the sensor placement ($d_k$). As the sensor moves, ray paths will stay unchanged except their final intersections with the sensor. For a $k$-element lens, there are at least $2k$ surfaces to trace through during a single call to $f()$, which shoots a large number of rays to compute the spotsize. Compared to this naive approach, caching these exit rays allows for a speedup of $2k\times$.

\vspace{4pt}
\textbf{Computational Cost.}
For each candidate lens system proposed by a lens splitting operation we must go through the continuous optimization process to measure its performance with optimized element spacing. For most lens systems with fewer than than 7 elements, the first stage of our continuous optimization (spotsize or OPD) takes less than 5 minutes. Run time is approximately proportional to the number of elements. 
The second stage of the continuous optimization (MTF optimization) takes approximately 20 minutes. This stage is relatively slow because the PSF's need to be rendered via wave optics for each field position in a single evaluation of the objective function. In practice, we only carry out MTF optimization for the most promising lens systems from the first stage of continuous optimization.
To evolve a design as described in Section \ref{sec:lenssplitting}, we allow a budget of 1200 CPU hours for each splitting iteration. This computation is distributed over a cluster of 600 nodes, one thread per node. This has proven to work well and the system can discover a good population of designs with similar performance.

\section{Experiments}
\label{sec:results}
We present three lens designs using \emph{Lens Factory}, covering a wide range of lens types. We also fabricate their corresponding prototypes for better evaluation. The first lens is micro 4/3 30mm f5.6. The performance of this lens shows that our system is capable of designing high performance lenses for more common applications. The second lens has non-parallel image and object planes, which is useful when the stand-off distance from the camera to the object plane is limited. 
The final example is a replacement optic for a virtual reality head mounted display (HMD). This lens forms a virtual image, rather than a real image as the micro 4/3 lens does. Virtual image optics are used when the system is designed to be viewed directly by the human eye.
We evaluate the discovered lenses using a combination of PSF visualization, MTF plots, and MTF50 values, which represent the maximum spatial frequency with $\text{MTF}\geq 0.5$. To further showcase our system, physical copies of these designs are fabricated by assembling parts ordered online with 3D printed housing, and evaluated against simulated results or the stock lens being replaced.

\begin{figure}[t]
\begin{center}
   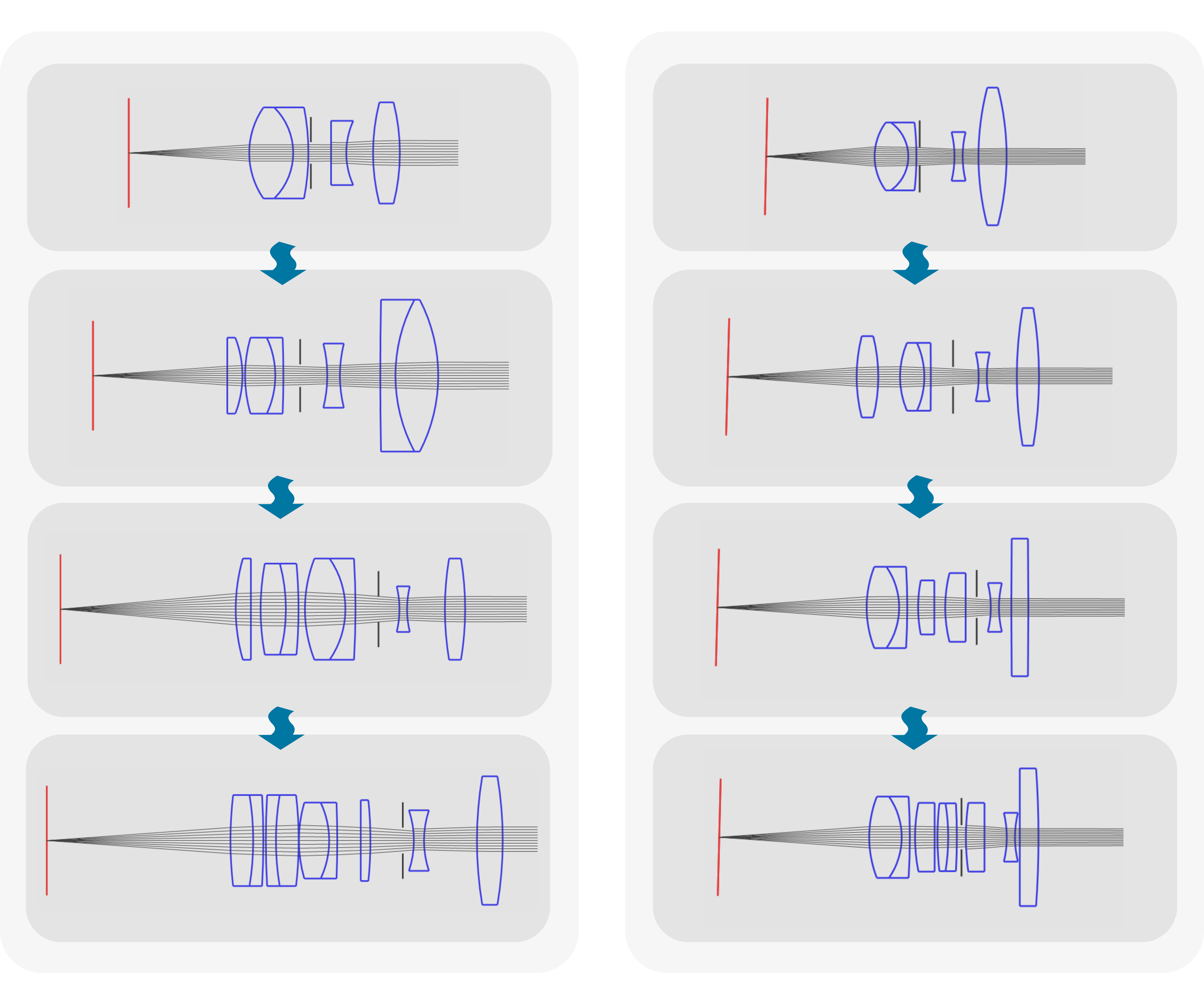
\end{center}
\vspace{-5pt}
\caption{Visualization of our lens design evolution during three rounds of lens splitting. Starting with a triplet design, we show the best performing lens after each splitting stage under the \emph{pool+swap} strategy. }
\label{fig:evolution}
\end{figure}

\subsection{Standard Micro Four Thirds Lens}
\label{sec:results_micro43}
First, we discuss our results on the micro 4/3 30mm f5.6 lens we have been using throughout previous discussions and illustrations. As shown in Figure \ref{fig:evolution}(a), we initialize the design with a triplet form and evolve to a 6-element lens via splitting.

\begin{figure}[t]
\begin{center}
   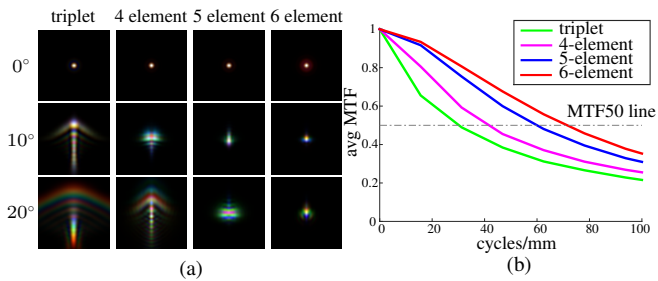
\end{center}
\vspace{-5pt}
\caption{Progression of simulated performance on the micro 4/3 standard lens. (a) PSF's at $0\degree$, $10\degree$ and $20\degree$ field angle for the best lens system discovered after each split using the \emph{pool+swap} strategy. Each PSF image is $65\mu m\times 65\mu m$ in size. (b) MTF evaluation for the best lenses, averaged across the field, color channels, and orientation (tangential/sagittal). Best view electronically.}
\label{fig:micro43simulated}
\end{figure}

Figure \ref{fig:micro43simulated} shows the progression of performance by visualizing the PSF's and MTF measurements. The PSF's for the best $k$-element designs for $k\in\{3,4,5,6\}$ at field angle $0\degree,10\degree,20\degree$ are shown in (Figure \ref{fig:micro43simulated}(a)). PSF's for RGB color channels are rendered separately and centered for visualization. A steady increase in average MTF performance is shown in Figure \ref{fig:micro43simulated}(b).

The starting triplet performs very well near the center but shows considerable astigmatism off-axis. Our discrete optimization is able to iteratively reduce such aberrations by splitting and introducing new lens elements into the system. As shown in Figure \ref{fig:micro43simulated}(b), the MTF50 response has more than doubled from the initial triplet design in three rounds of splitting. We further conduct the tolerance analysis (see Section \ref{sec:tolerance}) to pick the best 6-element design, and summarize the ideal vs. expected MTF50 performance in Table \ref{tab:micro43table}.

\begin{table}[t]
\vspace{5pt}
\tbl{MTF50 Performance for the Standard Micro Four Thirds Lens}{
\centering
\begin{tabular}{l|l|l|l|l|l|l|}
      & \multicolumn{3}{c|}{Ideal (LW/PH)} & \multicolumn{3}{c|}{Expected (LW/PH)}\\ \hline
      & $0\degree$      & $10\degree$       & $20\degree$  & $0\degree$      & $10\degree$       & $20\degree$    \\ \hline
red   & 2652 & 1794 & 1170   & 962 & 910  & 546 \\ \hline
green & 3380 & 2132 & 1326    & 1066  & 988 & 598\\ \hline 
blue  & 2626 & 2522 & 1508    & 936 &  858 & 494 \\ \hline

\end{tabular}}
\vspace{5pt}
\label{tab:micro43table}
Per channel MTF50 performance (LW/PH) for the best 6-element standard micro 4/3 lens at three different field angles, averaged over orientation (sagittal and tangential). Both the ideal performance and expected mean performance from tolerance analysis are shown.
\end{table}

\begin{figure}[t]
\begin{center}
   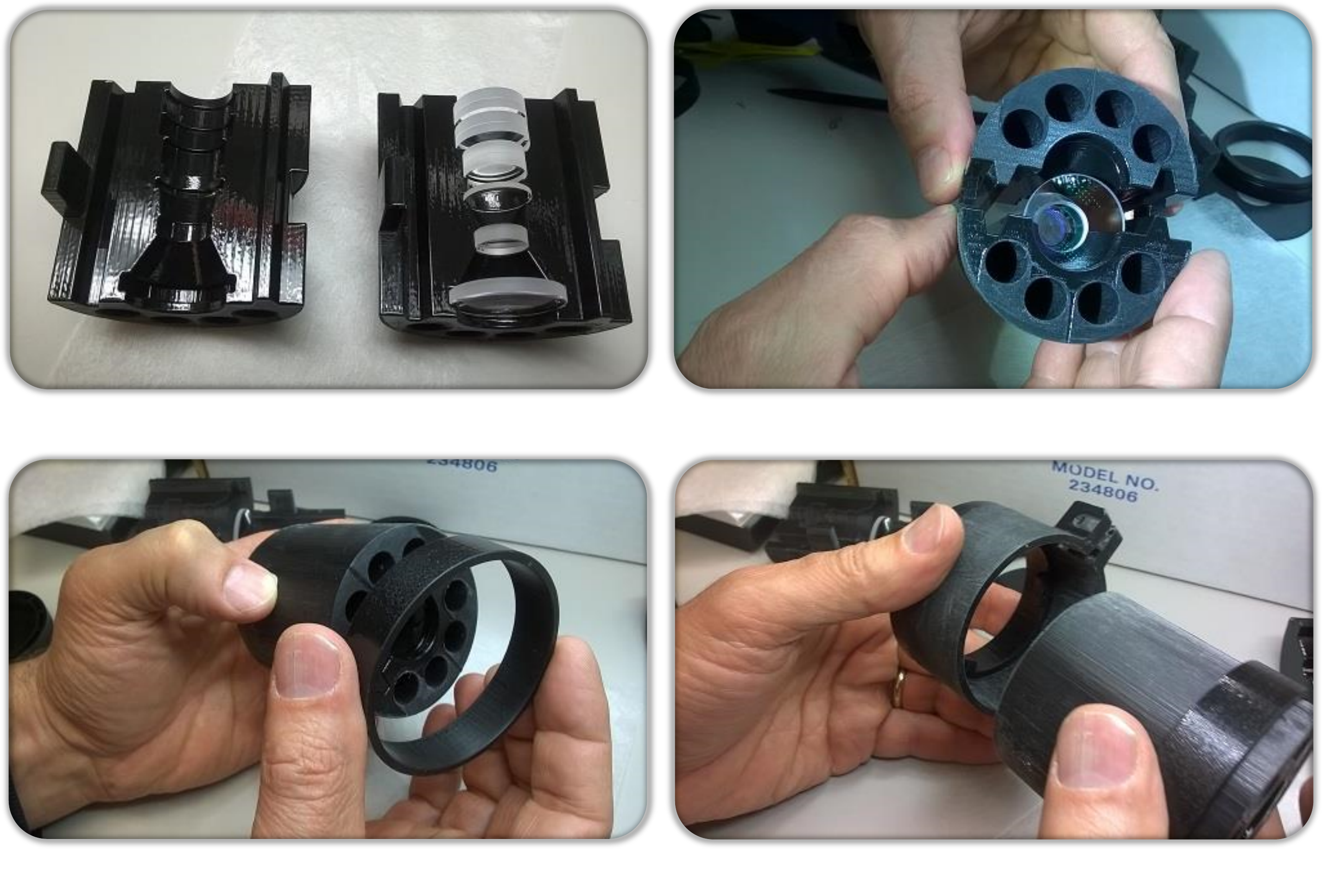
\end{center}
\vspace{-5pt}
\caption{ Our 3D printed lens housing takes the form of two clamshells (a). When the clamshells are closed (b), the lens elements are snapped in place by small crush ribs. Finally, we install the retaining ring (c) and the focusing sleeve (d).}
\label{fig:housing}
\end{figure}

\begin{figure*}[t]
\begin{center}
   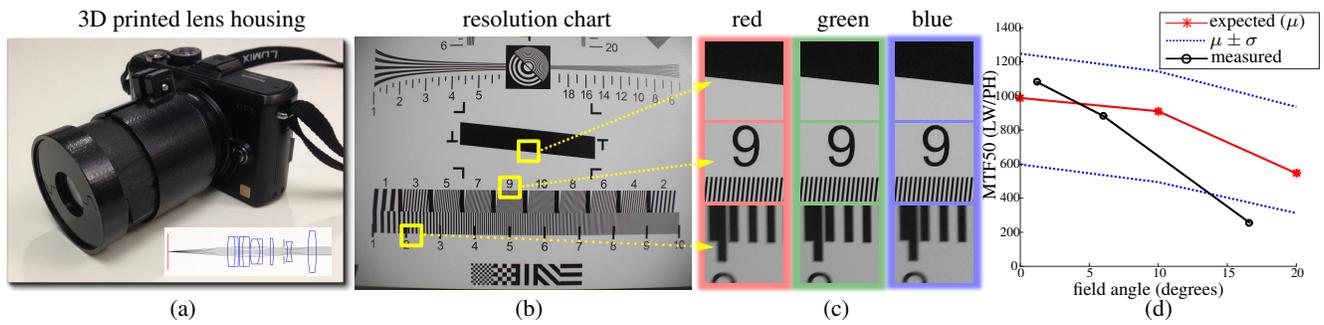
\end{center}
\vspace{-8pt}
\caption{(a) The actual lens we built for the 6-element standard lens (design shown in inset), mounted on a Panasonic GF1 camera body. (b) Resolution chart image taken by the lens and camera, without any post-processing. (c) Close-up view for image details. (d) Expected MTF50 performance range after tolerance analysis, compared against the performance measured from (b). Measured results are within the predicted performance bounds. Best viewed electronically.}
\label{fig:standard_prototype}
\end{figure*}

\begin{figure*}[t]
\begin{center}
   \input{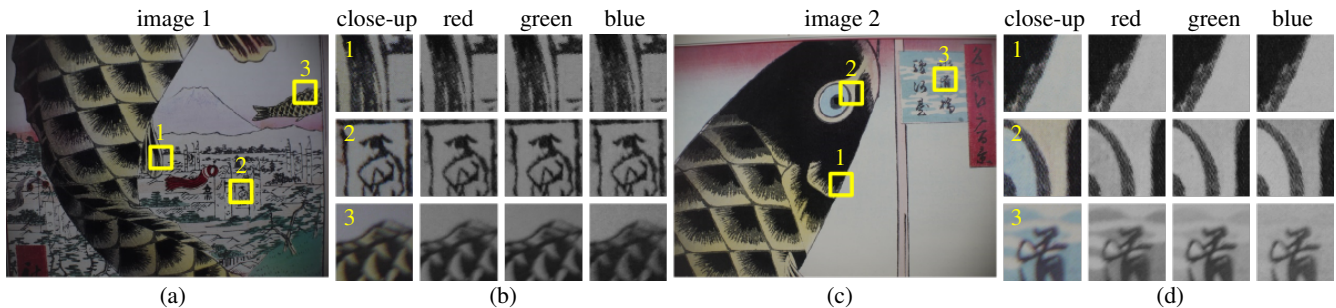}
\end{center}
\vspace{-8pt}
\caption{(a) and (c): Two uncorrected photos taken by our fabricated standard micro 4/3 lens (Figure \ref{fig:standard_prototype}(a)). We show three close-up crops in (b) and (d). Observe the fine textures on the surface of the artwork seen through the fabricated lens. Each color channel appears sharp. Best viewed electronically.}
\label{fig:naturalimages}
\end{figure*}

\begin{figure*}[!t]
\begin{center}
   \input{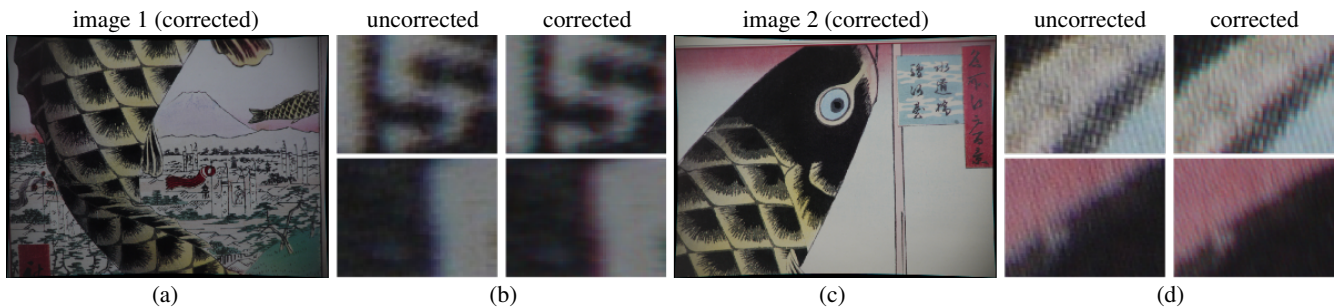}
\end{center}
\vspace{-8pt}
\caption{(a) and (c): images in Figure~\ref{fig:naturalimages} corrected for geometric distortion and chromatic aberration. Notice the edges of the white board appears straight in (c) after our correction step. Comparison of image details are shown in (b) and (d). Best viewed electronically.}
\label{fig:corrected}
\end{figure*}

\vspace{4pt}
\textbf{Fabricating the Standard Lens Prototype}. We built a prototype for our 6-element standard micro 4/3 lens (Figure \ref{fig:standard_prototype}(a)) and evaluated its imaging performance using a Panasonic GF1 camera body. The stock numbers of required lens parts are reported by our system and then ordered online. 

The 3D housing can be made in several ways. The CAD models can be downloaded directly from the vendor websites and individually imported into a CAD program along with the air gaps from the Lens Factory file. We have found it more convenient to generate a Zemax file containing the complete lens design specs and then export a CAD file. This is read into a CAD program and boolean subtracted from a generic lens housing tube made of two interlocking clamshells, shown in Figure \ref{fig:housing}.

We printed the housing with an Objet Eden 260 using Vero Black material. Because current 3D printers are not precise enough to exactly match the lens dimensions we generate a 0.02mm offset surface around each lens element before doing the boolean subtraction. This provides enough tolerance so that the lens slots will never be too small for the elements. 

To prevent the lens elements from rattling in the housing, small crush ribs are added to the CAD model around the circumference of each element. When the clamshell is closed, these ribs partially collapse and exert a constant force on the elements, holding them in place. A 3D printed retaining ring (Figure \ref{fig:housing}(b)) is slid over the mated clamshells to hold them to together. Finally, the assembly is inserted into a focusing sleeve (Figure \ref{fig:housing}(d)) that allows the inner lens housing to move in and out.

The complete process from setting up the user input spec (Section \ref{sec:UI}) to having a fully assembled lens takes less than a week, including shipping from vendors. Figure \ref{fig:standard_prototype} shows the actual lens mounted on the camera, a resolution chart image taken through the lens, as well as a comparison of the measured MTF50 response vs. the expected performance range from our tolerance analysis. The measured performance is well within our predicted performance bounds (Figure \ref{fig:standard_prototype}(d)). In addition, we show two natural images taken by this prototype in Figure \ref{fig:naturalimages}. The textured details on the artwork can be clearly seen in the images taken through our lens. Additional post-processing results for correcting geometric distortion and chromatic aberration are presented in Fig\ref{fig:corrected}.

It is observed that corner performance drops faster than expected (Figure \ref{fig:standard_prototype}(d)). This is caused by two factors: (1) the lens holders in the housing block rays within 1mm from the lens edge hence deteriorating corner sharpness, (2) we strive to achieve best perceived center sharpness when mounting the lens system onto the camera body, whereas our optimization would typically suggest a sensor placement that maximizes the average sharpness across the field. Still, we find that our fabricated lens outperforms the LUMIX G 14-45mm f/3.5-5.6 aspherical kit lens that came with the GF1 camera in sharpness (e.g., at the same focal length and f-number MTF50 of 1082 vs. 951 LW/PH) for most of field of view, except for the extreme corners.

\begin{figure}[t]
\begin{center}
   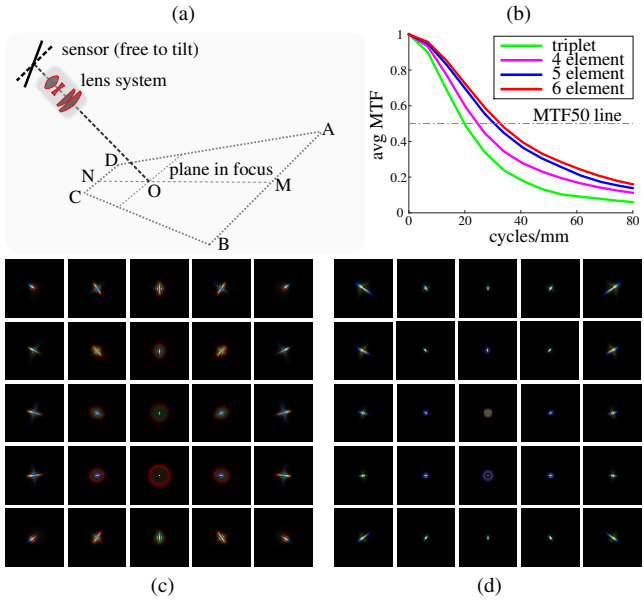
\end{center}
\vspace{-10pt}
\caption{(a) A visualization of the view camera setup. Notice the angled lens system in relation to the object plane and the image plane. (b) MTF evaluation for the best lenses, averaged across the field, color channels, and orientation (tangential/sagittal). We show PSF's for a $5\times 5$ rectangular grid over the object plane for the best triplet lens (c) and the best 6-element lens (d) discovered under the \emph{pool+swap} strategy. Each PSF image is $130\mu m\times 130\mu m$ in size.   }
\label{fig:viewcamerasimulated}
\end{figure}
\subsection{Non-parallel Projection: View Camera Lens}
\label{sec:viewcamera}
A view camera is a well known type of camera with a flexible bellows that holds the lenses, which allows complex movements such as tilt, shift, swing, \etc. We consider a simplified view camera application in Figure \ref{fig:viewcamerasimulated}(a) which requires a custom mounted lens system that is tilted relative to the image plane (sensor), in order to focus on an object plane not perpendicular to the optical axis. We assume that the lens is to be mounted on the same Panasonic GF1 micro 4/3 camera body, and that tilt is the only movement required to bring the object plane in focus.

One example in which such a lens might be useful is gesture recognition and hand tracking where the user interacts with a display via gestures and hand movements near the display (the display does not need to be touch sensitive), but the camera system can only be mounted on the side to avoid interrupting the user, hence creating a non-parallel projection between the object plane and image plane (sensor). Since the lens plane is not parallel to the object plane, the Scheimpflug principle states that the sensor need to tilt in the opposite direction of the object plane in order to make points on the object plane better in focus.

To model the sharpness of the entire object plane, we sample a regular grid of emitter positions to cover the left half of the object plane (the trapezoid AMND in Figure~\ref{fig:viewcamerasimulated}(a)). Performance on the other half is the same due to symmetry. Our discrete and continuous optimization procedures are carried out to optimize for the imaged sharpness for these emitter positions only. In particular, we set up the system in Figure~\ref{fig:viewcamerasimulated}(a) with OM=55cm, ON=25cm, CN=ND=25cm, BM=MA=65cm, and the object plane is 65cm away tilted at $45\degree$.

Admittedly, this design is more inherently more challenging due to its non-traditional requirements. In traditional lens design, this might amount to adopting drastically different design techniques than those used for designing a standard lens, or incorporating expert knowledge for such custom designs. However, the whole procedure stays unchanged for our Lens Factory system, and all the design and optimization challenges are entirely hidden from the user. The user only has to initialize the input specs through the UI, clicking a few more times than what is required for the standard lens setup. An example user interaction for this setting up this design is provided in our supplementary video.

We show the lens design progression in Figure \ref{fig:evolution}(c) and the simulated performance in Figure \ref{fig:viewcamerasimulated}. Again, our system finds highly improved designs within three rounds of iteration, almost doubling the average MTF performance of the starting triplet design (see Figure~\ref{fig:viewcamerasimulated}(c)). Comparing the rendered PSF's going from Figure \ref{fig:viewcamerasimulated}(c) to (d), one sees that the center PSF's become much more peaked, indicating higher center performance.

\begin{figure*}[t]
\begin{center}
   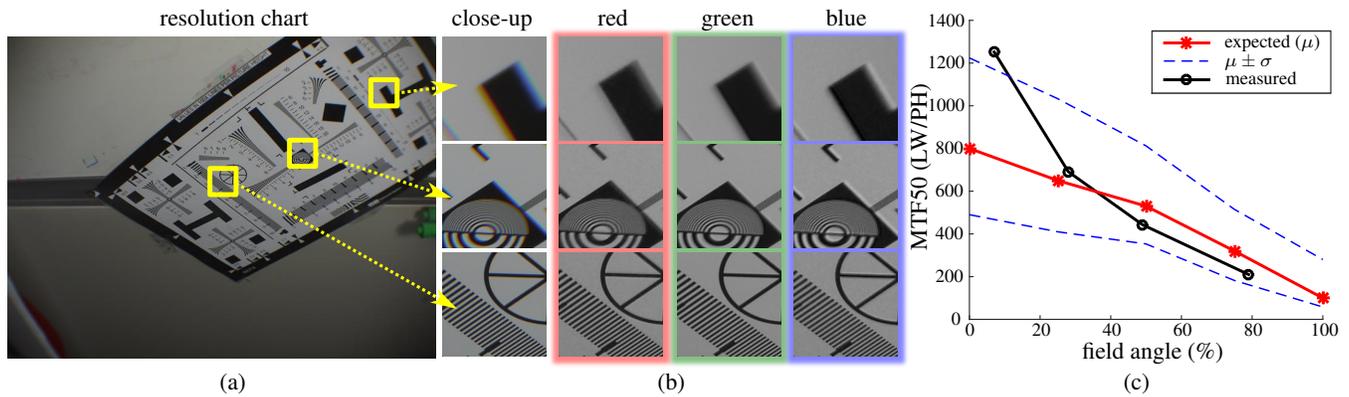
\end{center}
\vspace{-5pt}
\caption{(a) Resolution chart image on a tilted object plane. (b) Close-up image crops to showcase per-channel sharpness at various field angles. (c) The measured performance compares favorably against expected performance from tolerance analysis.}
\label{fig:viewcamera_measurements}
\end{figure*}

Similar to Section~\ref{sec:results_micro43}, we conduct tolerance analysis on this design, and build a lens prototype for the best performing 6-element lens with 3D printed housing. As shown in Figure ~\ref{fig:viewcamera_measurements}(c), the measured center performance is well above 1000LW/PH. Performance falls within expectation towards the corners. However, we also see that the measured performance drops at a faster rate compared to the expected MTF50 curve. This can be attributed to the same set of reasons discussed in Section~\ref{sec:results_micro43}.

While Figure ~\ref{fig:viewcamera_measurements}(a) shows a moderate amount of chromatic aberration off-center, it can be seen in Figure ~\ref{fig:viewcamera_measurements}(b) that each color channel remains relative sharp, so a post-processing step could be carried out as done for the standard lens in Figure~\ref{fig:corrected} if desired.

\begin{figure}[t]
\begin{center}
   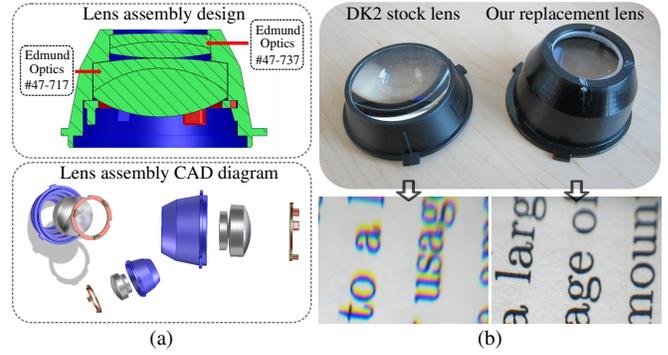
\end{center}
\vspace{-5pt}
\caption{(a) Our 2-element lens design with CAD diagrams for the lens assembly. (b) Top: visual comparison of the Oculus Rift DK2 virtual reality headset stock lens (left) and our HMD replacement lens (right). The stock lens is a single element plastic lens which exhibits severe chromatic and spherical aberration, especially outside a central field of view of about $30\degree$. Bottom: images taken through the Oculus Rift stock lens (left) and our replacement lens (right), looking at the very edge of the field of view. Our lens has much better image quality with minimal chromatic aberration.}
\label{fig:HMDprototype}
\end{figure}

\begin{figure*}[t]
\begin{center}
   \input{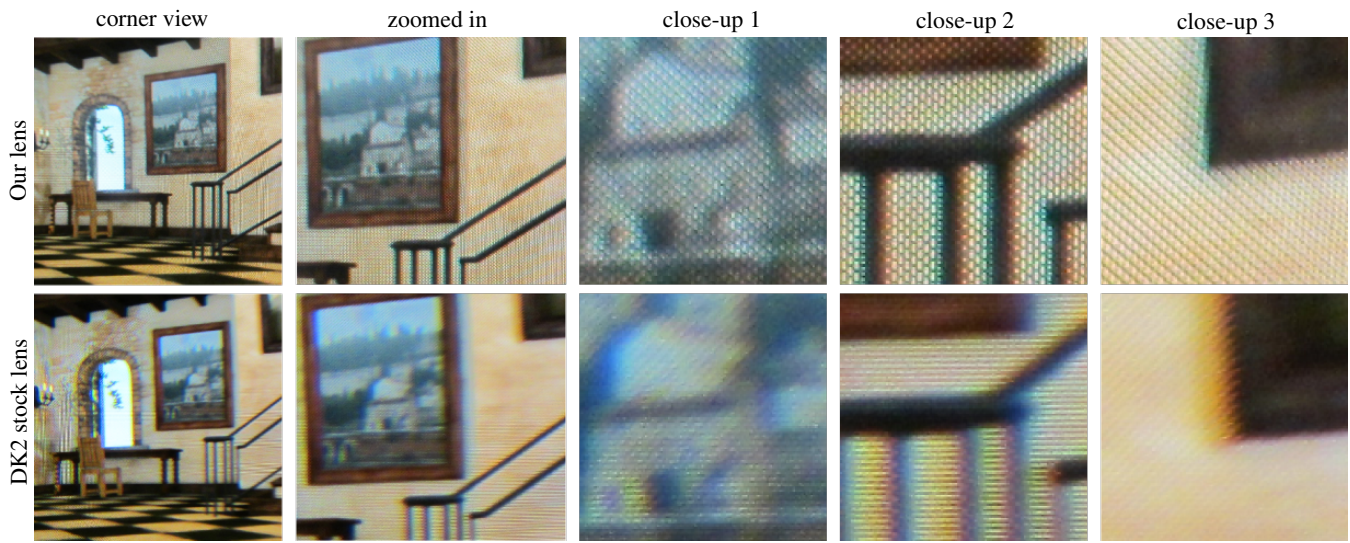}
\end{center}
\caption{This is a view of the Oculus Rift display as seen by our lens (top row) and the Oculus Rift lens (bottom row). The Oculus lens has significant residual chromatic aberration, even after the software pre-processing that Oculus performs to correct chromatic aberration. For our lens chromatic aberration is so small that no software correction is needed. Other aberrations are also much smaller. Best viewed electronically.}
\label{fig:hmdcornerview}
\end{figure*}

\subsection{HMD Lens System}
Virtual reality (VR) is a field where optics play a vital role. A recent success story is the Oculus Rift, which sports a custom molded plastic lens that enables a large FOV by creating a virtual image of a flat panel display. Since this is a single element lens it exhibits significant aberrations across the field. Undoubtedly this design was chosen to minimize cost and weight.

However, in our VR lab users rarely wear the headset for more than 30 minutes so weight is not a concern and neither is cost. Instead we wanted a design that maximized visual quality across the field. We designed an improved 2-element lens using off-the-shelf components discovered by our Lens Factory system. 

The Oculus Rift DK2 has several physical constraints we entered into the system specification: the lens system cannot exceed 60mm, there should be 10mm clearance between the user's cornea and the last surface of the lens system, and the desired FOV is approximately $100\degree$.

We model the human visual system as an ideal camera, with an ideal lens (cornea) 10mm behind the last surface of the lens system. The design is initialized as a single positive element and a brute force search is carried out over all the positive elements in the catalog. A total of 40 possible lens elements satisfy the physical constraints. 

A 2-element system is evolved by splitting the single element candidates. In Figure \ref{fig:HMDprototype}(b) we show a prototype of the 2-element lens we have built to retrofit the Oculus Rift DK2 and its imaging quality. The stock lens is on the left and the new lens is on the right. Underneath the lenses is a picture of the extreme edge of the field of view. The new lens has much superior sharpness, remaining clear almost to the very edge. Chromatic aberration is almost completely eliminated. The new lens gives the visual impression of a much higher resolution display.

Figure \ref{fig:hmdcornerview} shows the Oculus Rift display as seen by the our lens (top row) and the stock lens (bottom row). The Oculus software does image pre-processing to correct for chromatic aberration before displaying it, but a substantial amount remains. Our lens is noticeably sharper and essentially free of chromatic aberration so no software correction is necessary. 

To better facilitate VR development and enable reproducible research, we have provided public online access to our design details, CAD files and source code\footnote{\url{http://research.microsoft.com/en-us/um/redmond/projects/lensfactory/oculus/}} necessary to build and use this replacement lens for the Oculus Rift HMD. The source code works with the Unity game engine to correct the lens distortion for the Oculus display. In addition, we have fabricated multiple copies of this replacement lens and distributed them among several groups in the community with great feedback and success.

\section{Discussion and Limitations}

We believe \textit{Lens Factory} is the first tool to allow non-experts to create complex, high quality, and inexpensive lens systems from scratch. There are, however, several limitations. The most significant one is that the available off-the-shelf lens components were not designed with a system like \emph{Lens Factory} in mind. They sample the space of lens elements very coarsely and cannot be combined in simple ways to generate intermediate sample points. In spite of this, the performance of the lens systems we have built so far have been satisfactory and rewarding.

It should be possible to design a set of lens elements that do a much better job of sampling the design space. For example, lenses could be designed in powers of 2 so any desired power could be closely approximated with a few elements. More meniscus elements would also greatly improve the performance of wide angle systems in particular.

A less significant limitation is that current 3D printers are relatively imprecise compared to the tolerances required for optical systems. Our systems have acceptable performance for our applications but performance could be much better with tighter tolerances. This has not been a problem for the systems we have made so far but could be an issue for lower f\# designs. However, the user always has the option of milling out more precise housing parts using CNC machines if required.

Finally, significant computation is required for each design. We are actively investigating better pruning heuristics to reduce computation.


\bibliographystyle{acmtog}
\bibliography{lensfactory}

\begin{thebibliography}{}

\bibitem[\protect\citeauthoryear{Bates}{Bates}{2010}]{BatesMTF:doi:10.1117/12.868932}
{\sc Bates, R.} 2010.
\newblock Thru focus mtf optimization in lens design.

\bibitem[\protect\citeauthoryear{Brady, Gehm, Stack, Marks, Kittle, Golish,
  Vera, and Feller}{Brady et~al\mbox{.}}{2012}]{Brady:2012}
{\sc Brady, D.~J.}, {\sc Gehm, M.~E.}, {\sc Stack, R.~A.}, {\sc Marks, D.~L.},
  {\sc Kittle, D.~S.}, {\sc Golish, D.~R.}, {\sc Vera, E.~M.}, {\sc and} {\sc
  Feller, S.~D.} 2012.
\newblock {Multiscale gigapixel photography}.
\newblock {\em Nature\/}~{\em 486,\/}~7403 (June), 386--389.

\bibitem[\protect\citeauthoryear{Cheng, Xu, Wang, and Wang}{Cheng
  et~al\mbox{.}}{2014}]{RapidStock:doi:10.1117/12.2075390}
{\sc Cheng, D.}, {\sc Xu, C.}, {\sc Wang, Q.}, {\sc and} {\sc Wang, Y.} 2014.
\newblock Rapid lens design and prototype with stock lenses.
\newblock {\em Proc. SPIE International Optical Design Conference 2014\/}~{\em
  9293}.

\bibitem[\protect\citeauthoryear{Cheng, Wang, Hao, and Sasian}{Cheng
  et~al\mbox{.}}{2003}]{AutoAdditionCheng:03}
{\sc Cheng, X.}, {\sc Wang, Y.}, {\sc Hao, Q.}, {\sc and} {\sc Sasian, J.}
  2003.
\newblock Automatic element addition and deletion in lens optimization.
\newblock {\em Appl. Opt.\/}~{\em 42,\/}~7 (Mar), 1309--1317.

\bibitem[\protect\citeauthoryear{Cossairt, Miau, and Nayar}{Cossairt
  et~al\mbox{.}}{2011}]{scalingLawForSphericalLenses}
{\sc Cossairt, O.}, {\sc Miau, D.}, {\sc and} {\sc Nayar, S.} 2011.
\newblock {A} {S}caling {L}aw for {C}omputational {I}maging {U}sing {S}pherical
  {O}ptics.
\newblock {\em OSA Journal of Optical Society America\/}.

\bibitem[\protect\citeauthoryear{Cossairt and Nayar}{Cossairt and
  Nayar}{2010}]{Cossairt:2010}
{\sc Cossairt, O.} {\sc and} {\sc Nayar, S.} 2010.
\newblock Spectral focal sweep: Extended depth of field from chromatic
  aberrations.
\newblock In {\em Computational Photography (ICCP), 2010 IEEE International
  Conference on}. 1--8.

\bibitem[\protect\citeauthoryear{Levin, Fergus, Durand, and Freeman}{Levin
  et~al\mbox{.}}{2007}]{Levin:2007:IDC:1275808.1276464}
{\sc Levin, A.}, {\sc Fergus, R.}, {\sc Durand, F.}, {\sc and} {\sc Freeman,
  W.~T.} 2007.
\newblock Image and depth from a conventional camera with a coded aperture.
\newblock In {\em ACM SIGGRAPH 2007 Papers}. SIGGRAPH '07. ACM, New York, NY,
  USA.

\bibitem[\protect\citeauthoryear{Levoy, Ng, Adams, Footer, and Horowitz}{Levoy
  et~al\mbox{.}}{2006}]{Levoy:2006:LFM}
{\sc Levoy, M.}, {\sc Ng, R.}, {\sc Adams, A.}, {\sc Footer, M.}, {\sc and}
  {\sc Horowitz, M.} 2006.
\newblock Light field microscopy.
\newblock In {\em ACM SIGGRAPH 2006 Papers}. SIGGRAPH '06. ACM, New York, NY,
  USA, 924--934.

\bibitem[\protect\citeauthoryear{Manakov, Restrepo, Klehm, Heged\"{u}s,
  Eisemann, Seidel, and Ihrke}{Manakov et~al\mbox{.}}{2013}]{Manakov:2013:RCA}
{\sc Manakov, A.}, {\sc Restrepo, J.~F.}, {\sc Klehm, O.}, {\sc Heged\"{u}s,
  R.}, {\sc Eisemann, E.}, {\sc Seidel, H.-P.}, {\sc and} {\sc Ihrke, I.} 2013.
\newblock A reconfigurable camera add-on for high dynamic range, multispectral,
  polarization, and light-field imaging.
\newblock {\em ACM Trans. Graph.\/}~{\em 32,\/}~4 (July), 47:1--47:14.

\bibitem[\protect\citeauthoryear{Pamplona, Mohan, Oliveira, and
  Raskar}{Pamplona et~al\mbox{.}}{2010}]{Pamplona:2010:NID}
{\sc Pamplona, V.~F.}, {\sc Mohan, A.}, {\sc Oliveira, M.~M.}, {\sc and} {\sc
  Raskar, R.} 2010.
\newblock Netra: Interactive display for estimating refractive errors and focal
  range.
\newblock In {\em ACM SIGGRAPH 2010 Papers}. SIGGRAPH '10. ACM, New York, NY,
  USA, 77:1--77:8.

\bibitem[\protect\citeauthoryear{Sasian and Descour}{Sasian and
  Descour}{1998}]{Sasian_Descour}
{\sc Sasian, J.~M.} {\sc and} {\sc Descour, M.~R.} 1998.
\newblock Power distribution and symmetry in lens system.
\newblock {\em Optical Engineering\/}~{\em 37}, 1001–1004.

\bibitem[\protect\citeauthoryear{Shih, Guenter, and Joshi}{Shih
  et~al\mbox{.}}{2012}]{lensfitting_eccv2012}
{\sc Shih, Y.}, {\sc Guenter, B.}, {\sc and} {\sc Joshi, N.} 2012.
\newblock Image enhancement using calibrated lens simulations.
\newblock In {\em ECCV}. 42--56.

\bibitem[\protect\citeauthoryear{Smith}{Smith}{2000}]{smith_MOE}
{\sc Smith, W.} 2000.
\newblock {\em Modern Optical Engineering}.

\bibitem[\protect\citeauthoryear{Smith}{Smith}{2004}]{smith_lensdesign}
{\sc Smith, W.} 2004.
\newblock {\em Modern Lens Design}.

\bibitem[\protect\citeauthoryear{Traub, Hoffmann, Hengesbach, and Loosen}{Traub
  et~al\mbox{.}}{2014}]{StockLasers:doi:10.1117/12.2074508}
{\sc Traub, M.}, {\sc Hoffmann, D.}, {\sc Hengesbach, S.}, {\sc and} {\sc
  Loosen, P.} 2014.
\newblock Automatic design of multi-lens optical systems based on stock lenses
  for high power lasers.

\bibitem[\protect\citeauthoryear{Wilburn, Joshi, Vaish, Talvala, Antunez,
  Barth, Adams, Horowitz, and Levoy}{Wilburn
  et~al\mbox{.}}{2005}]{Wilburn:2005:HPI}
{\sc Wilburn, B.}, {\sc Joshi, N.}, {\sc Vaish, V.}, {\sc Talvala, E.-V.}, {\sc
  Antunez, E.}, {\sc Barth, A.}, {\sc Adams, A.}, {\sc Horowitz, M.}, {\sc and}
  {\sc Levoy, M.} 2005.
\newblock High performance imaging using large camera arrays.
\newblock {\em ACM Trans. Graph.\/}~{\em 24,\/}~3 (July), 765--776.

\bibitem[\protect\citeauthoryear{Zhou, Miau, and Nayar}{Zhou
  et~al\mbox{.}}{2012}]{ZhouFocalSweep}
{\sc Zhou, C.}, {\sc Miau, D.}, {\sc and} {\sc Nayar, S.} 2012.
\newblock {F}ocal {S}weep {C}amera for {S}pace-{T}ime {R}efocusing.
\newblock Tech. rep. Nov.

\bibitem[\protect\citeauthoryear{Zhou and Nayar}{Zhou and
  Nayar}{2011}]{computationalCameras}
{\sc Zhou, C.} {\sc and} {\sc Nayar, S.} 2011.
\newblock {C}omputational {C}ameras: {C}onvergence of {O}ptics and
  {P}rocessing.
\newblock {\em IEEE Transactions on Image Processing\/}~{\em 20,\/}~12 (Dec),
  3322--3340.

\end{thebibliography}


\end{document}